\documentclass[conference]{ieeeconf}  

\IEEEoverridecommandlockouts                              
                                                          
\usepackage[latin1]{inputenc}
\usepackage[T1]{fontenc}

\usepackage{graphicx} 
\graphicspath{{pics/}}
\usepackage{color}
\usepackage{mathptmx} 
\usepackage{times} 
\usepackage{amsmath} 
\usepackage{amssymb}  

\usepackage{flafter}

\title{\LARGE \bf
Pushing the limits of the \cybergrasp for haptic rendering
}

\author{Manuel Aiple and André Schiele
\thanks{Accepted to ICRA2013, doi:10.1109/ICRA.2013.6631073. Manuel Aiple is with the European Space Agency,
2201 AZ Noordwijk, The Netherlands.
        {\tt\small manuel.aiple at esa.int}.
André Schiele is with the European Space Agency,
2201 AZ Noordwijk, The Netherlands, and with the Delft University of Technology,
2628 CD Delft, The Netherlands. {\tt\small andre.schiele at esa.int}}}

\newcommand{\figurepriority}{tpb}
\newcommand{\figureprioritydouble}{b}

\newcommand{\cybergrasp}{CyberGrasp\textsuperscript{\texttrademark}}
\newcommand{\cyberglove}{CyberGlove\textsuperscript{\textregistered}}
\newcommand{\cyberglovesys}{CyberGlove Systems LLC}

\begin{document}

\maketitle
\thispagestyle{empty}
\pagestyle{empty}

\begin{abstract}
The \cybergrasp is a well known dataglove-exoskeleton device combination that
allows to render haptic feedback to the human fingers. 
Its design, however, restricts its usability for teleoperation through a limited
control bandwidth and position sensor resolution. Therefore the system is
restricted to low achievable contact stiffness and feedback gain magnitudes in
haptic rendering. Moreover, the system prohibits simple adaption of its
controller implementation.

In this paper, the ExHand Box is
presented, a newly designed back-end to widen the \cybergrasp's bandwidth
restrictions and to open it up for fully customized controller implementations.
The ExHand Box provides a new computer, interface electronics and motor
controllers for the otherwise unmodified \cyberglove and \cybergrasp hand
systems.
The loop frequency of the new system can be freely varied up to 2 kHz and
custom controllers can be implemented through an automatic code generation
interface.

System performance identification experiments are presented that demonstrate
improved behavior in hard contact situations over a range of sampling periods.
Maximum contact stiffnesses of up to 50kN/m in a stable condition are
demonstrated, which is significantly higher than what could be achieved with
the non-customized original system version.

Moreover, a bilateral control experiment is conducted to demonstrate the new
system's usability for generic teleoperation research. In this experiment a
raycasting algorithm is introduced for pre-contact detection in order to
compensate for high delay and jitter communication links between master and
slave as they appear in an Ethernet network. It is demonstrated that the
contact stiffness can be maintained in the order of magnitude of the system
performance identification with a demonstrated stiffness of 41kN/m in a stable
condition.
\end{abstract}

\section{INTRODUCTION}

   \begin{figure}[tp]
      \centering
      \includegraphics[width=0.95\columnwidth]{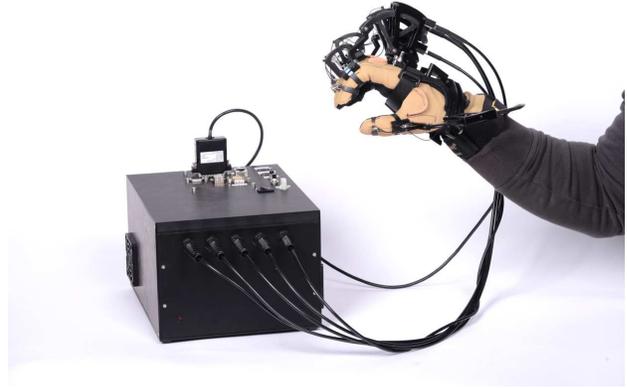}
      \caption{The ExHand Box with \cybergrasp and \cyberglove connected. The
      only external interface required is power through a laptop-like supply.}
      \label{fig:cybergraspbox}
   \end{figure}

The \cybergrasp, a product of \cyberglovesys, is a hand-grounded force-feedback
device to provide haptic feedback to the fingers of the hand.
It is designed to be used together with the separately available \cyberglove
dataglove that provides joint angle information of the hand and fingers.
This combination of the two devices has been used in research in different
domains, e.g. in haptic rehabilitation \cite{zhou2009}, in training of visually
impaired \cite{tzovaras2004} and in industrial training applications
\cite{hosseini2002}.
Within the Telerobotics \& Haptics Laboratory at the European Space Agency
(ESA), the CyberGrasp system has been used for studies on user performance
depending on visual- and force-feedback quality during teleoperated grasping
\cite{lii2010}.

In order to extend the commercial form of the \cybergrasp for optimal usability
in teleoperation research, following improvements are important:
\begin{itemize}
\item Increase the control bandwidth by at least a factor of ten.
\item Add flexibility in controller prototyping.
\item Reduce the amount of hardware (the commercial version requires 
a desktop-PC sized controller computer, the
\cyberglove electronics interface unit, the \cybergrasp motor box).
\end{itemize}

It is well known that time discretization, bandwidth limitations as well as
position and other sensor quantization effects limit the maximum stable
feedback gains that can be achieved with a given haptic device
 \cite{diolaiti2006} \cite{abbott2005}.
Especially the sampling rate limitation of the \cybergrasp system at 90 Hz is a
limiting factor for achievable high wall stiffnesses in rendering scenarios and
for overall telerobotics implementations.
While the position sensor quantization is adequate, with a 1000 pulse per
revolution encoder, no gearing is implemented between the motor and the cable
outputs of the \cybergrasp.
In total, this severely restricts the achievable contact stiffness of the
system in haptic applications in virtual or real contact environments.
At the same time, the mechanical implementation of the \cybergrasp incorporates
cable transmissions, which can have positive effects on stabilizing an haptic
system through small amounts of naturally occurring viscous and Coulomb
friction (e.g. damping).

Therefore we initiated the new design of the back-end electrical and computing
interface of the \cybergrasp system, which resulted in the
development of the ExHand Box (Fig. \ref{fig:cybergraspbox}).

The new ExHand Box was designed with following requirements:
\begin{itemize}
\item Achievable control loop frequency more than 1kHz.
\item Controller programmable via MATLAB/Simulink.
\item Small hardware configuration with one table-top box including all required subsystems.
\end{itemize}

It is the goal of this paper to present the design of the new ExHand Box system
for teleoperation research.
A system performance identification experiment will be carried out to quantify
the achievable feedback characteristics of the new system.
Experimentally achievable feedback loop frequencies and their maximum feedback
stiffness/damping gains will be determined during stable and passive
interaction with virtual walls.
Furthermore, a practical example for its use in a bilateral control application
with a dexterous robot hand simulation will be given.

\section{EXHAND BOX SYSTEM OVERVIEW}
In the process of redesigning the \cybergrasp all electronics were stripped off
and replaced. The only components remaining of the original system are the
dataglove and the exoskeleton with the motors (without motor drives).

\begin{figure}[bp]
      \centering
      \includegraphics[width=0.9\columnwidth]{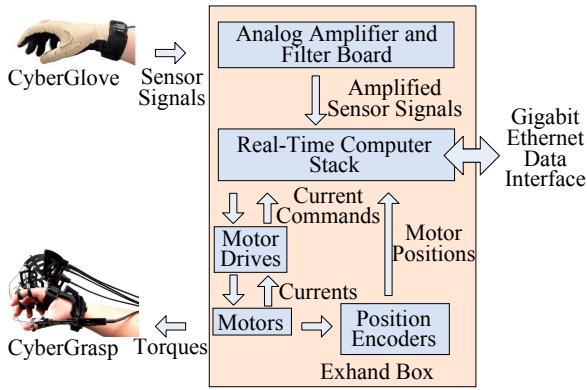}
      \caption{Overview of the ExHand Box system. The \cyberglove sensor input
      is amplified and filtered by the ExHand Box and torques are provided to the \cybergrasp
      for haptic feedback. The system can communicate with a slave device via a
      Gigabit Ethernet link.}
      \label{fig:systemprinciple}
   \end{figure}

Figure \ref{fig:systemprinciple} shows an overview of the new system. All
signals between the dataglove and the controller and between the controller and
the exoskeleton are sampled at the step size of the model running on the
controller which can be configured up to 2kHz. The current control loop of
the new motor drives runs at 4kHz.

The controller is a PC with a 1.6GHz Intel Atom processor according to the
PC/104 standard running the xPC Target real-time operating system.

The sensor information from the 22 sensors in the dataglove is amplified and
filtered by 22 independent amplifier and filter chains. A Diamond MM-32X-AT
Analog I/O board is used to sample the amplified sensor signal with a resolution
of 0.3mV (22bit resolution over the amplified signal voltage range). Its
multiplexed maximum sample rate is 250kHz.

The motor position encoders are 1000 pulse per turn quadrature encoders and
their signals are decoded using two Sensoray S526 boards. These also have four
D/A channels and eight A/D channels, some of which are used to control the
analog servo motor drives and to monitor the motor currents. The motor drives
used are miniature Elmo Castanet with a maximum continuous current of 3.3A and
a maximum peak current of 6.6A.
Table \ref{tab:controllerresolutions} summarizes the relevant measured resolutions of
the ExHand controller. The estimated quantization is conservative, since the peak-to-peak
noise level of the actual signals has been used. The signals are stable (i.e. not flipping) to 
the specified number of bits.
 
\begin{table}[tp]
\caption{Signal resolutions of the ExHand controller}
\label{tab:controllerresolutions}
\begin{center}
\begin{tabular}{|c|c|c|c|}
\hline
Signal & Noise Level & Resolution \\
& (Peak-to-Peak)& (Stable Bits)\\
\hline
Joint Angle & 6.1mV & 11\\
Motor Position & 1 encoder step & 9\\
Current & 47.2mA & 7 \\
\hline
\end{tabular}
\end{center}
\end{table}

The controller and the motors are powered with 24VDC. The data connection to the
controller is implemented through Gigabit Ethernet.

The complete box with controller, motors, motor drives and analog electronics
measures 250mm x 210mm x 155mm with a mass of 4kg and is entirely self sustained.

\section{SYSTEM STEP RESPONSE IDENTIFICATION}
\subsection{Experimental Setup}
\begin{figure}[bp]
      \centering
      \includegraphics[height=4.2cm]{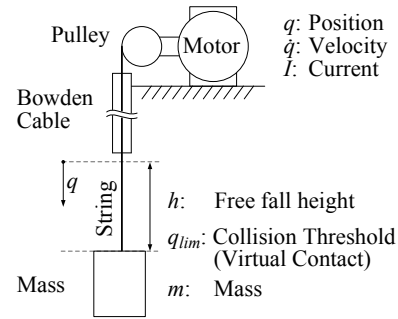}
      \caption{Schematic of the experimental setup. The mass is dropped from
      a rest position and stopped by the controller when reaching the virtual
      contact. The virtual contact is defined by the motor position which
      corresponds to the free fall height.}
      \label{fig:experimentsetup}
   \end{figure}

In order to identify the step response characteristics of the system at
different gain settings an experiment was carried out which consisted in
letting a mass $m$ fall from a given height $h$ and make the system stop its fall
(Fig. \ref{fig:experimentsetup}).
This tests the response of the system for collision rendering at high speed. It
is important to note here, that the force output of the system can not be
measured directly through a sensor, therefore calibrated masses have been used
to determine an exact force input step.

The mass used for this experiment was 500g and the free fall height 7cm.
   
\begin{figure*}[\figureprioritydouble]
      \centering
      \includegraphics[width=0.73\textwidth]{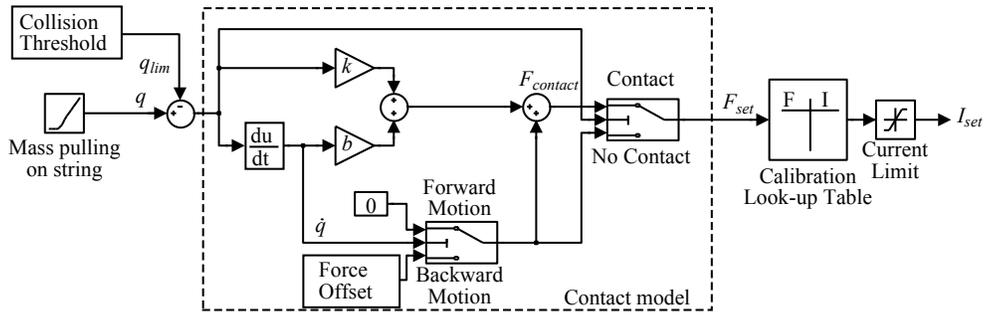}
      \caption{Representation of the controller model used for the step response
      performance identification.
      $q$: motor position;
      $\dot{q}$: motor velocity;
      $q_{lim}$: virtual contact position;
      $k$: virtual spring constant;
      $b$: virtual damper constant;
      $F_{contact}$: calculated contact force;
      $F_{set}$: force command;
      $I_{set}$: current command;
      }
      \label{fig:identificationmodel}
   \end{figure*}

The controller model used for this experiment is depicted in figure
\ref{fig:identificationmodel}.

The model uses the motor encoder information to determine if the virtual
contact condition applies or not. This is the case if the position reaches the
collision threshold which corresponds to the free fall height $h$.
In this case the force command $F_{contact}$ as calculated according to equation
\ref{eq:identificationequation} applies.

\begin{equation}
F_{contact} = b \dot{q} + k q + F_0
\label{eq:identificationequation}
\end{equation}

With $b$ being the damping factor (in Ns/m), $k$ the spring constant (in N/m)
and $q$ the mass' position with respect to the collision threshold. $F_0$ is the
pullback force which is null if the string is pulled out (i.e. if $\dot{q} >=
0$) and large enough to overcome the friction in the Bowden cable and pull the
string back on the pulley if the string is released (i.e. if $\dot{q} < 0$).
This can be tuned with a ``Force Offset'' parameter.

If the contact condition does not apply, the force command $F$ equals $F_0$.
In the contact situation the model corresponds to a PD-Controller with the
proportional gain equal to $k$ and the derivative gain equal to $b$.
   
\subsection{Experimental Method}
Before any experiments, the system was calibrated to determine the conversion
table from desired force to the current command signal and the conversion table
from motor current monitoring signal to force on the string in the SI system.

The experiment was carried out for loop frequencies of 90Hz (corresponding to
the original system), 100Hz, 200Hz, 500Hz, 1000Hz and 2000Hz. Every test run
consisted of ten falls. After every run the parameters $k$ and $b$ were
increased until the system became unstable for the given frequency, then the
loop frequency was increased. The system was considered unstable when it
oscillated continuously after the virtual collision instead of coming to rest at
an end position.

During the experiments, the following data was logged on the controller with
the loop frequency at which the model was running:
\begin{itemize}
	\item experiment time $t$ (s)
	\item position $q$ (m)
	\item speed $\dot{q}$ (m/s)
	\item force command $F_{des}$ (N)
	\item current command $I_{des}$ (A)
	\item collision condition applying $c$ (true | false)
	\item force calculated from the motor current monitoring signal $F_{mon}$ (N)
\end{itemize}

The \emph{collision condition applying} signal was used to dissect the log data
of the ten falls into ten separate contact situations.

On the basis of the monitored force and position data of the contact situations
the stiffness was calculated. Therefore, the position and force data pairs of
all contact situations were taken together. Then the data pairs with a force
value out of the range of 0.8N to 4.5N were rejected in order to take into
account only points on the slope and remove the data points corresponding to the
time before letting the mass fall and after it was stopped by the motor.

The remaining pairs were sorted by increasing force. Then a ten values unweighted moving average was
calculated over the position values in order to reduce the effect of outliers
on the following linear regression.

Finally a linear regression was calculated over these new position and force
data pairs to get the stiffness as ratio of force over position.

\subsection{Results \& Discussion}
\begin{figure}[\figurepriority]
      \centering
      \includegraphics[width=0.9\columnwidth]{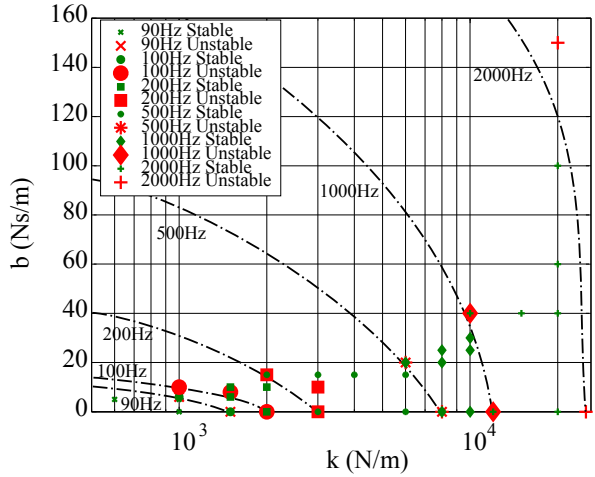}
      \caption{Stability of the system in the $k$-$b$-space for different loop
      frequencies. The dot-dashed lines are very roughly estimated stability borders
      at the respective loop frequency for indication only.}
      \label{fig:stabilityresults}
   \end{figure}

Figure \ref{fig:stabilityresults} shows the experimental stability results of
the system with different loop frequencies in the $k$-$b$-space. By increasing
the loop frequency to 2000Hz, the parameters $k$ and $b$ can be nearly 20 times
higher than with the original frequency of 90Hz.

Table \ref{tab:stiffnessmax} shows how this affects the achievable stiffness.
When comparing table \ref{tab:stiffnessmax} and figure
\ref{fig:stabilityresults}, attention should be paid to the fact that the
experiment with the highest stable parameters for the respective loop frequency
is not necessarily the one that gave the maximum stiffness as calculated from
the force over position curve.

\begin{figure}[\figurepriority]
      \centering
      \includegraphics[width=0.79\columnwidth]{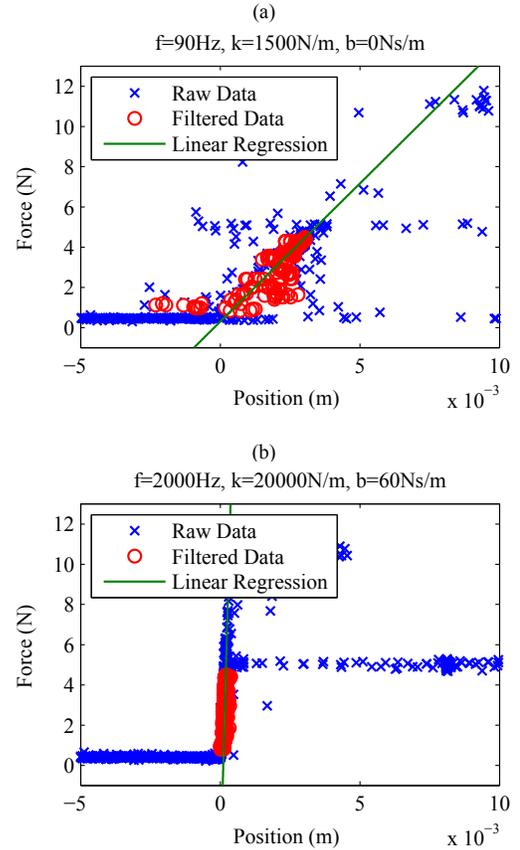}
      \caption{Force over Position plot from the performance identification,
      \emph{(a)} for $f=90\text{Hz}\:,\: k=1500\text{N/m}\:,\: b=0\text{Ns/m}$.
      The calculated stiffness is 1374N/m.
      \emph{(b)} for $f=2000\text{Hz}\:,\: k=20000\text{N/m}\:,\: b=60\text{Ns/m}$.
      The calculated stiffness is 50004N/m.
      Blue crosses: raw data, red circles: filtered data for regression,
      green line: linear fitting to the filtered data points.}
      \label{fig:fallexperimentcombined}
   \end{figure}

The maximum achieved stiffness of the virtual contact is 36 times higher with a
loop frequency of 2000Hz than with 90Hz (Fig. \ref{fig:fallexperimentcombined}).
This means that contacts can be rendered with higher crispness with the ExHand
Box controller than with the original \cybergrasp setup.

\begin{table}[tp]
\caption{Maximum achieved stiffness for different loop frequencies}
\label{tab:stiffnessmax}
\begin{center}
\begin{tabular}{|c|c|c|c|}
\hline
Frequency & k & b & Stiffness \\
(Hz)      & (N/m) & (Ns/m) & (N/m) \\
\hline
90 & 1500 & 0 & 1374 \\
100 & 2000 & 0 & 1506 \\
200 & 2000 & 15 & 7203 \\
500 &   6000 & 20 & 11977 \\
1000 &  10000 & 40 & 18936 \\
2000  & 20000 & 60 & 50004 \\
\hline
\end{tabular}
\end{center}
\end{table}

\section{BILATERAL CONTROL EXPERIMENT}
\subsection{Experimental Setup}
\begin{figure}[\figurepriority]
      \centering
      \includegraphics[width=0.82\columnwidth]{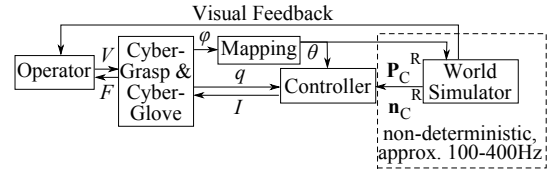}
      \caption{Principle of the bilateral control application.
      $V$: finger velocity;
      $F$: feedback forces;
      $\varphi$: glove joint angles;
      $\theta$: robot joint angles;
      $q$: motor positions;
      $I$: motor currents;
      $\mathbf{P}\mathrm{_C^R}$: collision plane point coordinates in the robot hand frame;
      $\mathbf{n}\mathrm{_C^R}$: collision plane normal vector;}
      \label{fig:bilateralprinciple}
   \end{figure}

A simple scenario of collision with virtual objects was chosen to test the
haptic rendering quality of virtual contacts with the ExHand Box. In this
scenario a person wearing the dataglove and exoskeleton should be able to
control a virtual hand and feel the contact with virtual objects. For this
purpose a World Simulator (virtual slave robot hand) program was written,
running on an independent computer connected to the ExHand Box (master
controller) via a Gigabit Ethernet connection (Fig. \ref{fig:bilateralprinciple}).

\subsection{Experimental Method}
\begin{figure}[\figurepriority]
      \centering
      \includegraphics[width=0.67\columnwidth]{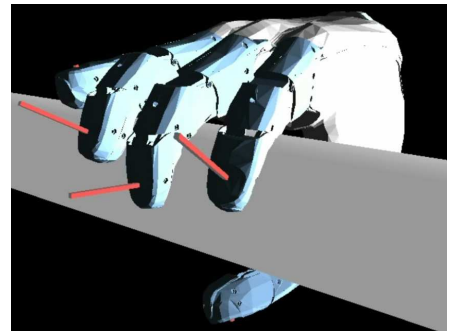}
      \caption{The visualization of the world simulator, the red bars show the
      contact forces.}
      \label{fig:worldsimulator}
   \end{figure}

The world simulator is responsible for the visualization of the virtual hand
and the virtual world (Fig. \ref{fig:worldsimulator}). Depending on the
complexity of the world model it renders the scene at a frame rate of
100-400fps. This frame rate corresponds to the frequency at which it receives
from and sends data to the ExHand controller.

It receives the joint angles for the virtual robot hand from the master
controller which feed a virtual model of the Hit Hand II from German
Aerospace Center (DLR).
The mapping of \cyberglove sensor signals to Hit Hand II joint angle values is
done on the ExHand controller using the mapping model developed for \cite{lii2010}.

Special care was taken to not negatively influence the performance of haptic
rendering by a non-deterministic bilateral link with variable delay and jitter.
This was achieved by keeping the collision detection completely on the
controller side and by introducing a pre-contact detection algorithm to change
the collision parameters on-line between master and slave.

\begin{figure}[\figurepriority]
      \centering
      \includegraphics[width=0.75\columnwidth]{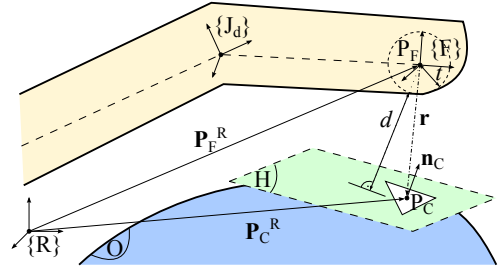}
      \caption{Pre-contact detection principle.
      \{R\}: robot hand frame;
      \{F\}: finger tip frame;
      \{$\mathrm{J_d}$\}: distal joint frame of the robot hand;
      O: collision object;
      $\mathrm{P_F}$: finger tip center point, origin of raycast;
      $\mathbf{r}$: direction of raycast;
      $\mathrm{P_C}$: collision point where the ray hits O;
      $\mathbf{n}_C$: normal vector of O at $\mathrm{P_C}$;
      H: collision plane defined by $\mathrm{P_C}$ and $\mathbf{n}_C$; 
      $d$: distance between $\mathrm{P_F}$ and H;
      $\mathbf{P}\mathrm{_F^R}$: coordinates of $\mathrm{P_F}$ in \{R\};
      $\mathbf{P}\mathrm{_C^R}$: coordinates of $\mathrm{P_C}$ in \{R\};
      $t$: radius of the fingerpad;
      }
      \label{fig:precontactdetection}
   \end{figure}

Figure \ref{fig:precontactdetection} shows a schematic view of this algorithm.
The World Simulator receives the robot finger joint angles $\theta$ from the
ExHand controller and calculates the position of the point $\mathrm{P_F}$ in the virtual
world, with $\mathrm{P_F}$ representing the point in the middle of the finger tip. It
then does a raycast from $\mathrm{P_F}$ in the direction $\mathbf{r}$. If the ray hits a
virtual collision object O, the coordinates of the collision point $\mathrm{P_C}$ in
the virtual world and the normal vector $\mathbf{n}_C$ of the surface of O at
the point $\mathrm{P_C}$ are determined. $\mathrm{P_C}$ and $\mathbf{n}_C$ define the collision
plane H. The world simulator translates the coordinates of $\mathrm{P_C}$ and
$\mathbf{n}_C$ into the robot hand frame, notated $\mathbf{P}\mathrm{_C^R}$ and
$\mathbf{n}\mathrm{_C^R}$ respectively.
It then sends $\mathbf{P}\mathrm{_C^R}$ and $\mathbf{n}\mathrm{_C^R}$ to the ExHand controller.

The ExHand controller also uses $\theta$ to calculate the robot finger tip
position $\mathbf{P}\mathrm{_F^R}$ in the robot hand frame \{R\}. It then calculates the distance
$d$, which is the point-plane-distance between $\mathrm{P_F}$ and H minus the radius of
the fingerpad $t$. $d$ is defined according to equation
\ref{eq:distancedefinition}, such that it is negative before the contact and
positive in the contact.

\begin{equation}
d = (\mathbf{P}\mathrm{_C^R}-\mathbf{P}\mathrm{_F^R})\cdot \mathbf{n}\mathrm{_C^R} - t
\label{eq:distancedefinition}
\end{equation}

\begin{figure*}[\figureprioritydouble]
      \centering
      \includegraphics[width=0.68\textwidth]{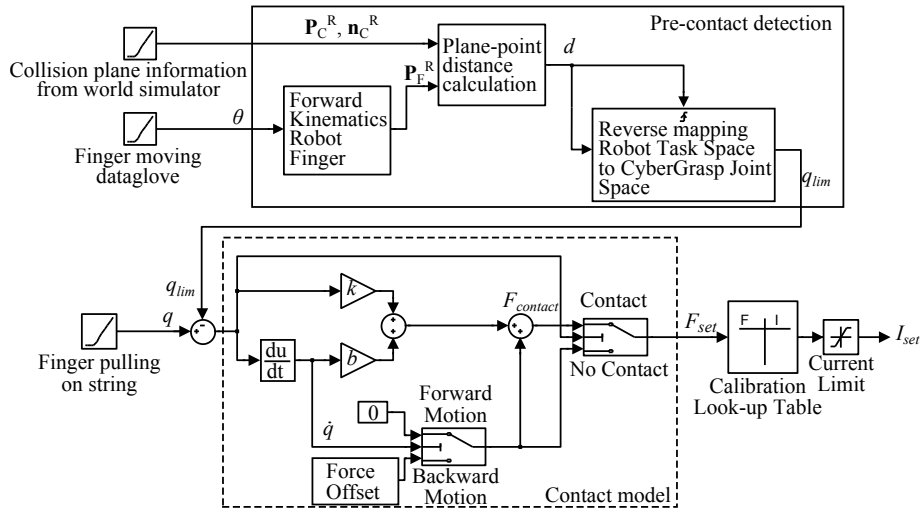}
      \caption{Representation of the controller model used for the bilateral
      control experiment.
      $q$: motor position;
      $\dot{q}$: motor velocity;
      $\theta$: robot joint angles;
      $\mathbf{P}\mathrm{_F^R}$: robot fingertip coordinates in robot hand frame;
      $\mathbf{P}\mathrm{_C^R}$: virtual collision plane point vector in robot hand frame;
      $\mathbf{n}\mathrm{_C^R}$: virtual collision plane normal vector in robot hand frame;
      $d$: distance between robot fingertip and virtual collision plane;
      $q_{lim}$: virtual contact position;
      $k$: virtual spring constant;
      $b$: virtual damper constant;
      $F_{contact}$: calculated contact force;
      $F_{set}$: force command;
      $I_{set}$: current command;}
      \label{fig:bilateralmodel}
   \end{figure*}

Figure \ref{fig:bilateralmodel} depicts the model running on the ExHand
controller.

The distance $d$ is used to do a reverse mapping from the robot task
space to the \cybergrasp joint space. It modifies $q_{lim}(t)$, the position
of the virtual contact $q_{lim}$ at a time $t$ according to the rule in equation
\ref{eq:freezecondition}.

\begin{equation}
q_{lim}(t) = 
\begin{cases}
    q_{lim}(t-dt)& \text{if } d\geq d_{lim}\\
    d              & \text{otherwise}\\
\end{cases}
\label{eq:freezecondition}
\end{equation}

Where $d_{lim}$ is a negative constant defining the threshold where the
controller should switch into contact mode and $dt$ is the simulation time step.
Thus $q_{lim}(t-dt)$ refers to the value of $q_{lim}$ in the previous iteration
of the simulation loop.

The effect of this rule is that once the robot fingertip is in a
critical distance to the virtual collision object, $q_{lim}$ is frozen and the
remaining way to the contact and the contact rendering itself is independent of
the world simulator. This effectively decouples the high frequency contact
rendering on the ExHand controller from the constraints of the non-deterministic bilateral
link with the world simulator.
   
As can be seen, the contact model itself is the same as in the step response
performance identification controller model.

Different system parameters were experimented with in the bilateral control
configuration, demonstrating higher perceived crispness with higher values of $k$ and $b$
as expected.

In the following, the experiment with the system parameters set to the values
identified in the system performance identification as the ones with the highest
stiffness will be discussed, i.e.
$f=2000\text{Hz}\:,\: k=20000\text{N/m}\:,\: b=60\text{Ns/m}$.

The test subject was asked to enter into contact situation several times at
different speeds: as slowly as possible, at what seemed to be a natural speed,
and as fast as possible.

The following signals were logged:
\begin{itemize}
	\item experiment time $t$ (s)
	\item position measured by the motor encoders $q$ (m)
	\item speed measured by the motor encoders $\dot{q}$ (m/s)
	\item force command $F_{des}$ (N)
	\item current command $I_{des}$ (A)
	\item collision condition applying $c$ (true | false)
	\item force calculated from the motor current monitoring signal $F_{mon}$ (N)
	\item distance to collision plane calculated from the dataglove input through
	the hand model and the collision plane information $d$ (m)
	\item collision threshold $q_{lim}$ (m)
\end{itemize}

\subsection{Results \& Discussion}
\begin{figure}[\figurepriority]
      \centering
      \includegraphics[width=0.68\columnwidth]{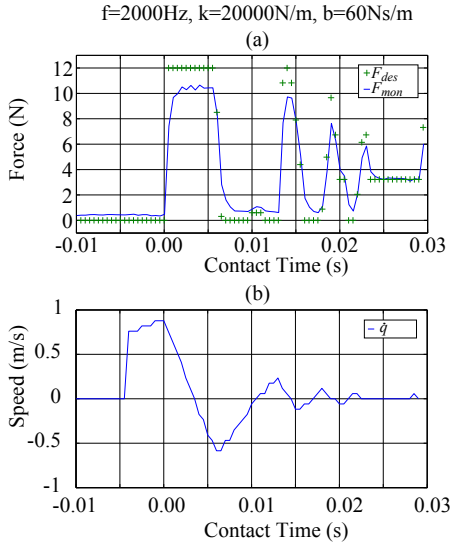}
      \caption{Time plot of a typical contact situation at 0.87m/s contact speed
      for $f=2000\text{Hz}\:,\: k=20000\text{N/m}\:,\: b=60\text{Ns/m}$.
      \emph{(a)} green crosses: force command $F_{des}$, blue line:
      force calculated from the motor current monitoring signal $F_{mon}$.
      The black vertical line shows the beginning of the contact.
      \emph{(b)} speed $\dot{q}$.
      }
      \label{fig:bilateralcontact}
   \end{figure}

Figure \ref{fig:bilateralcontact} shows a plot from a typical contact situation at
medium speed (0.87m/s when entering into the virtual contact). The system
response settles approximately 25ms after the first contact into a stable
equilibrum. For the test subject this transient was not noticeable, but
mechanical damping by the pulling string might have effects in this as well
which were not investigated further.

\begin{figure}[\figurepriority]
      \centering
      \includegraphics[width=0.8\columnwidth]{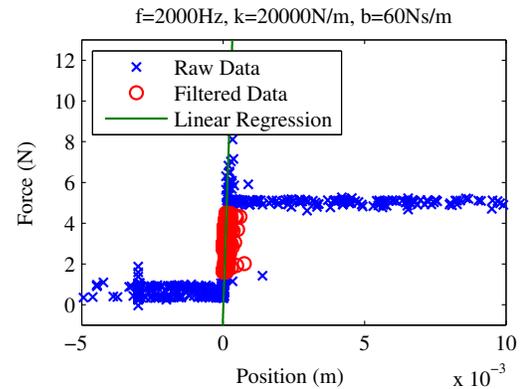}
      \caption{Force over Position plot from the bilateral control application
      for $f=2000\text{Hz}\:,\: k=20000\text{N/m}\:,\: b=60\text{Ns/m}$.
      The calculated stiffness is 41250N/m.
      Blue crosses: raw data,
      red circles: filtered data for regression,
      green line: linear fitting to the filtered data points.}
      \label{fig:bilateralstiffness}
   \end{figure}

Figure \ref{fig:bilateralstiffness} shows the force over position plot from a
run with the bilateral control experiment. It confirms the order of magnitude
of the stiffness values (the calculated stiffness for this run was 41250N/m)
from the step response performance identification also in bilateral control
applications, thus proving the positive effects of the pre-contact detection
mechanism.

\section{CONCLUSION}

It has been shown that using \cyberglove and \cybergrasp with the new ExHand Box
improves their performance significantly, with an increase of the stiffness of
up to 50kN/m, thus achieving a stiffness 36 times higher than before and enabling
crisper rendering of hard contacts.

The new controller of the ExHand Box can be easily customized in MATLAB/Simulink
with auto-code generation, achieving loop frequencies of up to 2kHz. Therefore,
it provides a platform to test different controller models.

A raycasting algorithm has been implemented successfully in a bilateral control
experiment for pre-contact detection to achieve similarly high performance of
41kN/m stiffness in a delay and jitter biased scenario where master and slave
device communicate via a typical non-dedicated Ethernet network.

%
%
%





\section*{ACKNOWLEDGMENT}
This research was supported in part by the German Aerospace Center (DLR).
Special thanks go to Neal Lii for providing the dataglove mapping model.


\end{document}